\documentclass[aps,pra,twocolumn,superscriptaddress]{revtex4}

\usepackage{color,graphicx}
\usepackage{epstopdf}
\usepackage{enumerate,verbatim,appendix}
\usepackage{amsmath,amssymb,amsthm,float,mathrsfs}

\newcommand{\del}{\vec{\nabla}}
\newcommand{\dy}{\overleftrightarrow}
\newcommand{\norm}[1]{\lvert{#1} \lvert}
\newcommand{\Rb}{^{87} \textrm{Rb}}
\newcommand{\nn}{\nonumber}

\newcommand{\inner}[2]{\langle {#1} \mid {#2} \rangle}
\newcommand{\proj}[1]{\lvert {#1} \rangle \langle {#1}  \lvert}
\newcommand{\ave}[1]{\langle {#1} \rangle}

\begin{document}
\title{Lenses as an Atom-Photon Interface: A Semiclassical Model}
\author{Colin Teo}
\affiliation{Centre for Quantum Technologies, National University of Singapore, Singapore}
\author{Valerio Scarani}
\affiliation{Centre for Quantum Technologies, National University of Singapore, Singapore}
\affiliation{Department of Physics, National University of Singapore, Singapore}

\begin{abstract}
Strong interaction between the light field and an atom is often achieved with cavities. Recent experiments have used a different configuration: a propagating light field is strongly focused using a system of lenses, the atom being supposed to sit at the focal position. In reality, this last condition holds only up to some approximation; in particular, at any finite temperature, the atom position fluctuates. We present a formalism that describes the focalized field and the atom sitting at an arbitrary position. As a first application, we show that thermal fluctuations do account for the extinction data reported in M. K. Tey \textit{et al.}, Nature Physics \textbf{4}, 924 (2008).
\end{abstract}

\maketitle

\section{Introduction}

Strong coupling between the light field and a single atom is a cornerstone of quantum optics. It has been achieved in a large number of laboratories around the world in the last decades \cite{Haroche_book,ye_quantum_2008,mabuchi_cavity_2002,ye_trapping_1999,hennrich_transition_2005,armani_ultra_2003}. The interface of choice is usually a high-finesse cavity: as well known, the mode density and field amplitude is increased when the cavity volume is decreased. Recently, some groups have started considering a simpler interface: \textit{lenses}. Focusing is in fact the natural way of concentrating light at a desired position, here that of the atom.

This choice is partly motivated by the perspectives in quantum information science. Complex information processing tasks will eventually require to scale up from the single quantum system to that of a network of systems. At each node, information could be processed on atoms or ions, whose states can be carefully manipulated and efficiently detected; to transmit information between the nodes, one would rather encode it in states of light. A crucial ingredient of such a network is the coherent operation of a matter-photon interface at the single-photon level \cite{kimble_quantum_2008}. As a first step towards this futuristic vision, strong coupling between atoms and \textit{propagating} light should be demonstrated. Experiments conducted on single molecules \cite{M strong extinction, M2 strong extinction} and single atoms \cite{tey_strong_2008, Syed strong 2009} have shown promising results.

A theoretical analysis of such a focusing configuration was done previously in \cite{van_enk_strongly_2001}, and rectified in \cite{tey_interfacing_2009}. However, in this latter paper, a significant discrepancy was reported between some experimentally measured data and the corresponding theoretical prediction. As usual, the theoretical model does not match the complexity of reality in many aspects. Here, we show that the discrepancy can be essentially accounted for if one includes the fact that the atom is at finite temperature and therefore is not sitting exactly at the position of the focus, but is rather wobbling around it. While this effect is pretty obvious, its description is not: one must be able to compute the electric field felt by the atom at any point close to the focus of the system. Due to the strong focusing, the paraxial approximation is not valid, so the well known closed form solutions for the electric field cannot be used.

In section \ref{sec:formalism} we develop the necessary formalism. We start by defining the propagation problem and solving it to find an integral equation for the electric field (\ref{sec:E_field}). This can later be solved numerically to obtain the field anywhere in the system. Further, we model the atomic response (\ref{ss:atomresp}) and the detection process (\ref{sec:detector}), with a special attention to the measurement of \textit{extinction} that was reported in \cite{tey_strong_2008,tey_interfacing_2009}. In section \ref{sec:temp}, we apply our formalism to quantify the effects of temperature. We show that we can fit the experimental data by assuming a temperature $T\approx 185\,\mu \mathrm{K}$ and discuss the meaning of this value.

\section{Formalism}
\label{sec:formalism}

\begin{figure*}[ht]
\centering
\includegraphics[width= 1.3 \columnwidth]{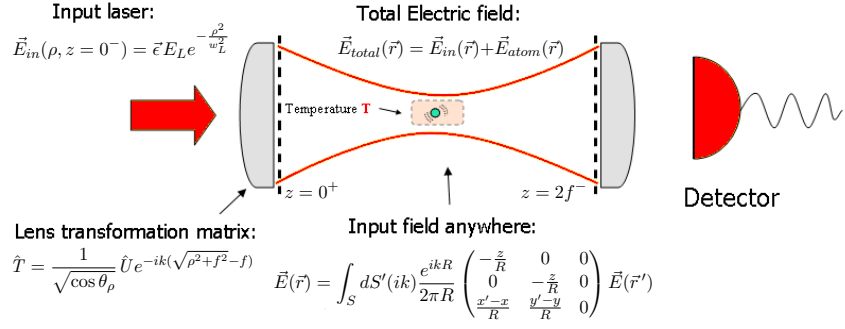}
\caption{Schematic view of the setup and the corresponding main equations. The atom is sitting somewhere near the focus of a confocal arrangement.} \label{figgeneral}
\end{figure*}

\subsection{Electric field}
\label{sec:E_field}
The electric field after it passes through the single mode fiber has, to a good approximation, the Gaussian shape
\begin{equation}\label{eqn:input}
  \vec{E}_{laser} (\rho,z=0^-)= \vec{\epsilon} \, E_L e^{-\frac{\rho^2}{w_L^2}},
\end{equation}
where we have used cylindrical coordinates $(\rho,\phi,z)$, $\vec{\epsilon}$ is the polarization of the field, $E_L$ is the amplitude, and $w_L$ represents the width of the beam.

We model the lens as the infinite plane $z=0$. The impinging electric field at $z=0^-$ is transformed according to
\begin{equation}
  \vec{E}(\rho,z=0^+) = \hat{\mathit{T}} \vec{E}(\rho,z=0^-)\,.
\end{equation}
The transformation matrix $\hat{\mathit{T}}$ has the following form:
\begin{equation}
\hat{\mathit{T}}=\frac{1}{\sqrt{\cos\theta_\rho}}\hat{\mathit{U}}(\rho,\phi)e^{-ik(\sqrt{\rho^2+f^2}-f)}, \label{eqn:trans}
\end{equation}
with $\cos\theta_\rho = \frac{z}{\sqrt{z^2+\rho^2}}$ and
\begin{eqnarray}
  \hat{\mathit{U}} (\rho, \phi) &=& \begin{pmatrix}
\cos\phi & -\sin\phi & 0 \\
\sin\phi & \cos\phi & 0 \\
0 & 0 & 1
  \end{pmatrix} \begin{pmatrix}
\cos\theta_\rho & 0 & -\sin\theta_\rho\\
0 & 1 & 0\\
\sin\theta_\rho & 0 & \cos\theta_\rho
\end{pmatrix} {} \nn \\
&& {} \begin{pmatrix}
\cos\phi & \sin\phi & 0 \\
-\sin\phi & \cos\phi & 0 \\
0 & 0 & 1
\end{pmatrix}
\end{eqnarray}
Notice that the factor of $\frac{1}{\sqrt{\cos\theta_\rho}}$ is required by energy flux conservation (cf. for instance Pg 391 of \cite{Griffiths_book}). Now, it remains to propagate the electric field in the confocal region. We do so with a Green's function approach: the electric field is a solution to the following equation,
\begin{equation}
  \del \times (\del \times \vec{E}) -k^2\vec{E} = 0.  \label{vHelm}
\end{equation}
Solving for the Dyadic Green's function defined as
\begin{equation}
  \del \times (\del \times \dy{G}) -k^2\dy{G} = \dy{1} \delta(\vec{r}-\vec{r}\,') \label{vHfG}
\end{equation}
we arrive at
\begin{equation} \label{eqn:propagate}
  \vec{E}(\vec{r}) = \int_S dS' (ik)\frac{e^{ikR}}{2\pi R}
  \begin{pmatrix}
    -\frac{z}{R} & 0 & 0 \\
    0 & -\frac{z}{R} & 0 \\
    \frac{x'-x}{R} & \frac{y'-y}{R} & 0
  \end{pmatrix}
  \vec{E}(\vec{r}\,'),
\end{equation}
where $S$ is the plane where we know fully the electric field, $R = \norm{\vec{r}-\vec{r}\,'}$, and  we have used the long wavelength approximation $kR << 1$. Details of the derivation can be found in Appendix \ref{sec:appelec}.

Although the integrand in this equation is highly oscillatory, we do not make a stationary phase approximation as it gives a form similar to the paraxial approximation, and this has been shown to give incorrect results \cite{tey_interfacing_2009}. We evaluate this integral numerically with usual Gaussian quadrature \cite{Numerics: book frm paper,Numerics: programming bible} and the method of Numerical Steepest Descent \cite{huybreachs_evaluation_2007}.

\subsection{Atomic response}
\label{ss:atomresp}

The $\Rb$ atom, modeled as a two-level system, is excited by a weak classical probe beam driving it at the resonant frequency $\omega$. The weak beam assumption allows us both to ignore incoherent scattering by the atom, and to write a fixed phase relation between the dipole and laser electric fields.

The assumption of resonance just means that, with respect to the phase of the field at the location of the atom, $\vec{r}_{a}$, the electric field produced by the atom should have a $\frac{\pi}{2}$ phase lag.

Combining these with the usual dipole electric field, we arrive at,
\begin{equation}
  \vec{E}_{a}(\vec{r})= \frac{3}{2k} \norm{\vec{\epsilon}\cdot \vec{E}_{L}(\vec{r}_{a})}(\dy{1}-\frac{\vec{r}\otimes\vec{r}}{r^2})\frac{\vec{\epsilon}}{r}
  e^{ikr}e^{i(\Phi_{L}-\frac{\pi}{2})} ,
\end{equation}
where $\vec{E}_{a}$ and $\vec{E}_L$ are the electric field of the atom and laser respectively, $\Phi_L-\frac{\pi}{2}$ represents the phase lag of the dipole and $\vec{\epsilon}$ is assumed to be the polarization of the laser field before the lens, and the dipole.

\subsection{Detectors}
\label{sec:detector}
In the experiments \cite{tey_strong_2008,tey_interfacing_2009}, the measured field is collected by a collection lens, coupled into a single-mode fiber, and is subsequently sent to a photon counting detector. Therefore, only the overlap between the collected electric field and the mode of the fiber will be ultimately detected. The calibration showed that essentially all the input light is detected at the output in the absence of an atom. In view of this, we assume that the mode of the fiber, $\vec{g}$, is proportional to the initial electric field at the $z=0^-$ plane, i.e.
\begin{eqnarray}
  \vec{g} &\propto& \vec{E}_{laser} (\rho,z=0^-) \nn \\
  &\propto& \vec{\epsilon}\, E_L e^{-\frac{\rho^2}{w_L^2}}\,.
\end{eqnarray}
The intensity recorded by the detector becomes
\begin{eqnarray}
  I_{det} &\propto& \norm{\inner{\vec{g}}{\vec{E}_{coll}}}^2 \nn \\
  &\propto& \norm{\inner{\vec{\epsilon}\, e^{-\frac{\rho^2}{w_L^2}}}{\vec{E}_{coll}}}^2
\end{eqnarray}
In the rest of this article, we will be concerned with the extinction, $\epsilon$, of the input field due to the presence of the atom defined as,
\begin{eqnarray}
\epsilon &=& \frac{I_{laser}-I_{system}}{I_{laser}}, \label{eqn:extinction} 
\end{eqnarray}
where $I_{laser}$ is the recorded intensity with no atom present in the setup, and $I_{system}$ is when a single atom is present.

\section{Finite temperature effects and comparison with experiment}
\label{sec:temp}
In the experiment, the atom is held in an approximately harmonic trap at a particular temperature. Thus, the two following major effects can be expected: Doppler shifts of the incoming laser field, and polarization mismatch due to strong focusing by the lenses.

\subsection{Doppler shifts}
To a good approximation, the potential of the atom in the dipole trap is harmonic and can be written,
\begin{equation}
  H = \frac{p^2}{2m} + \frac{1}{2}m\omega_\rho \rho^2 + \frac{1}{2}m\omega_z z^2\,,
\end{equation}
where $\rho^2=x^2+y^2$ and $\omega_\rho$ and $\omega_z$ are the trapping frequencies in the radial and longitudinal direction respectively.

We approximate the state of the atom as a thermal state
\begin{eqnarray}
\rho_{atom} = \frac{1}{Z} \sum_{n,m}\Pi_n \otimes \proj{m}_z  \, e^{-\frac{E^\rho_n+E^z_m}{k_B T}}  \label{eqn:atom_density}
\end{eqnarray}
here, $\proj{m}_z$ is the eigenstate of the harmonic oscillator in the z direction with eigenvalue $E_m^z = \hbar \omega_z (m+\frac{1}{2})$; while $\Pi_n$ is the projector on the eigenspace associated to the eigenvalue of $E^\rho_n = \hbar \omega_\rho (n + 1)$, so $\textrm{Tr}(\Pi_n) = n+1$.

Since the electric field is propagating in the $z$ direction, we need only calculate the rms velocity in this direction. The square of the momentum is,
\begin{eqnarray}
  \frac{\ave{p_z^2}}{m \hbar \omega_z} &=&  \Big(\textrm{Tr}(a^\dagger a \,\rho)+\frac{1}{2} \Big) \,=\, \Big(\frac{e^{-\alpha_z}}{1-e^{-\alpha_z}} + \frac{1}{2} \Big)
\end{eqnarray}
with $\alpha_z = \frac{\hbar \omega_z}{k_B T}$. The Doppler shift to lowest order is given by $kv$, where $k$ is the wavenumber, and $v$ is the velocity. Using an estimate of $100 \mu \textrm{K}$ for the temperature, we obtain a value of $kv \approx 800 \textrm{kHz}$. This doppler detuning is about 0.02 times the 30 MHz linewidth of the transition. This justifies us ignoring the Doppler shift and considering only the spatial averaging.

\subsection{Classical canonical ensemble}
Before we continue the averaging over the atomic spatial profile, it is worthwhile to reconsider in detail how the experiment is done. We know that the atom is initially loaded into the dipole trap from a MOT, and MOT clouds are typically about $0.1 \times T_{Doppler}$, which in our case is about $20 \mu \textrm{K}$.

This typical temperature is justification enough to use a fully quantum model to describe the experiment. However, rough estimates of the heating due to finite data collection time put the temperature in the $100 \mu \textrm{K}$ range, in which the simpler classical treatment is justified. Therefore, for any function $f(\vec{r}_{atom},\vec{p}_{atom})$ which depends on the position and momentum coordinates of the atom, the average value will be given by
\begin{eqnarray}
  \ave{f(\vec{r}_{atom},\vec{p}_{atom})} &=& \textrm{Tr} \Big( f(\vec{r}_{atom},\vec{p}_{atom}) \rho_{atom}\Big), \nn \\
  &\approx& \frac{1}{Z} \int e^{-\frac{\mathscr{H}}{k_B T}} f(\vec{r}_{atom},\vec{p}_{atom}) d^3p d^3r, \nn \\
  &\approx& \frac{1}{Z} \int e^{-\frac{V(\vec{r}_{atom})}{k_B T}} f(\vec{r}_{atom}) d^3r, \label{eqn:func ave}
\end{eqnarray}
where we have used the a classical canonical ensemble to describe the system and $\mathscr{H} = \frac{p^2}{2m} + V(\vec{r}_{atom})$ is the classical Hamiltonian describing the system.

\subsection{Comparison with experiment}
\label{sec:results}
The extinction defined in \eqref{eqn:extinction} can be reduced to,
\begin{equation}
  \epsilon = 1-{\Big \lvert 1+\frac{\inner{\vec{E}_{a}}{\vec{E}_{L}}}{\inner{\vec{E}_{L}}{\vec{E}_{L}}} \Big \lvert}^2, \label{eqn:epsilon}
\end{equation}
where $\vec{E}_a$ and $\vec{E}_L$ are the electric fields of the atom and laser respectively, on the plane of detection. Since the propagation of the field is unitary, and we assume infinite detectors, we can move the collection plane to that of the second lens in the confocal arrangement, $z=2f$. This means that,
\begin{equation}
  \epsilon(x_a,z_a) = 1- \left| 1-\frac{3}{w_L^2}\tilde{I}^*(x_a,z_a) \tilde{K}(x_a,z_a) \right|^2,
\end{equation}
is the extinction of the probe beam in the atom were at $(x_a, 0, z_a)$, where we have taken the atom to be in the x-z plane, and $\tilde{I} (x_{a},z_{a})$ and $\tilde{K} (x_{a},z_{a})$ are shown in Appendix \ref{sec:app_ext}. Notice that due to cylindrical symmetry of the system, it is enough to consider the atom at $(x_a, 0, z_a)$.
Finally, the measured extinction must be the averaged extinction, weighted in the canonical ensemble as in \eqref{eqn:func ave}. So we have to compute,
\begin{eqnarray}
  \ave{\epsilon(\vec{r}_{atom})} &=& \frac{1}{Z} \int e^{-\frac{V(\vec{r}_{atom})}{k_B T}} \epsilon(\vec{r}_{atom}) d^3r \nn \\
  &=&  \frac{1}{Z} \int \rho \, d\rho \, dz\, e^{-\frac{m(\omega_\rho^2 \rho^2 + \omega_z^2 z^2)}{2k_B T}} \epsilon(\rho,z).  \label{eqn:ave_extinct}
\end{eqnarray}
We first plot extinction vs temperature for a particular value of the focusing parameter, $u=\frac{w_L}{f}$, where $w_L$ represents the width of the initial gaussian beam, and $f$ is the focal length of the lens used.
\begin{figure}[H]
\centering
\includegraphics[width=\columnwidth]{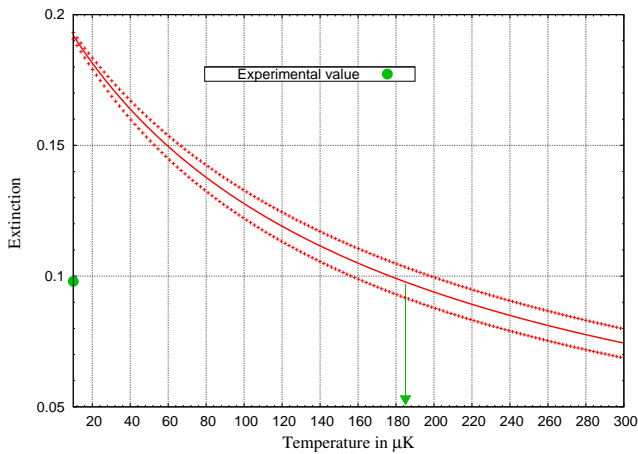}
\caption{The above is a plot of extinction vs temperature for a focusing parameter $u=0.29$, $\omega_\rho = (2\pi) 56 \pm 4 \, \textrm{kHz} $ and $\omega_z = (2\pi) 7 \pm 0.25 \, \textrm{kHz}$ as used in the experiments. The dotted lines are the effects of the experimental uncertainties in the trap parameters. This gives a prediction of the temperature of the atom to be about 185 $\pm$ $20 \mu \textrm{K}$.} \label{fig:ex vs T}
\end{figure}

Now, following reference \cite{tey_interfacing_2009}, we plot curves of extinction vs focusing strength, including the effects of a finite temperature of the atom, for 10, 50 and $185 \mu \textrm{K}$.
\begin{figure}[H]
\centering
\includegraphics[width=\columnwidth]{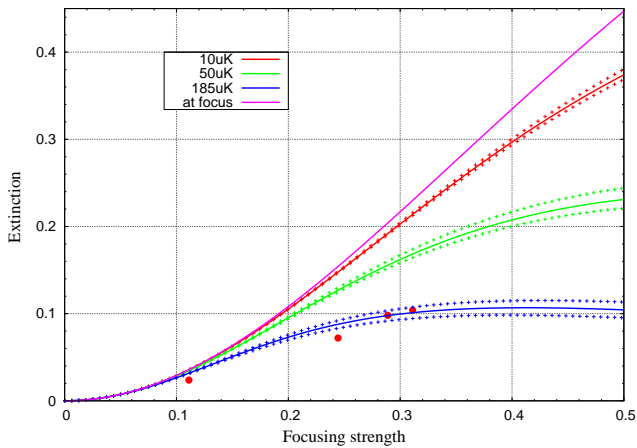}
\caption{(Color online) Plot of extinction vs focusing. The curve representing the theoretical maximum for an atom stationary at the focus (i.e. $T=0$), as well as the experimental points (filled circles), are the same as in \cite{tey_interfacing_2009}. The other curves are plotted for for $T=$10, 50 and $185 \mu \textrm{K}$ respectively. The dotted lines come from the experimental uncertainty in the trapping frequencies.} \label{fig:ex vs u zoomed}
\end{figure}
Figure \ref{fig:ex vs u zoomed} shows that the experimental point seem to follow the correct trend if we just include a higher temperature. This shows that the temperature must be the main cause of the discrepancy between theory and experiment. Now we plot for a wider range of focusing strengths to show the expected trend for a larger focusing.
\begin{figure}[H]
\centering
\includegraphics[width=\columnwidth]{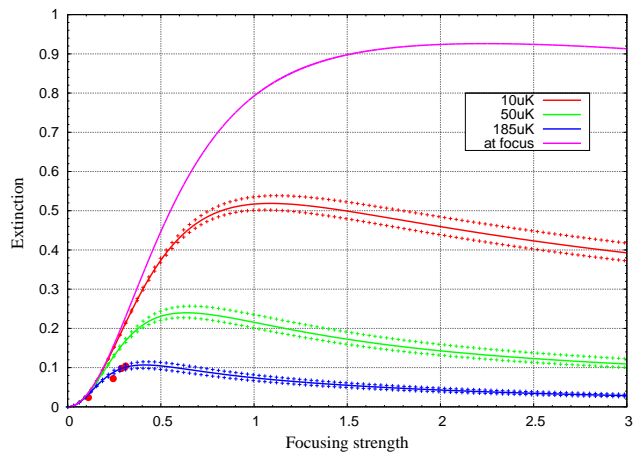}
\caption{(Color online) Plot of extinction vs focusing: same as in Fig.~\ref{fig:ex vs u zoomed}, but for a larger range of focusing parameters.} \label{fig:ex vs u }
\end{figure}

\subsection{The meaning of the temperature}
\label{sec:meaning}
The previous section shows that, from the simulations, we should be expecting temperatures of around $180 \mu \textrm{K}$, which is higher than the Doppler cooling limit for $\Rb$. This requires further investigation, since we are expecting cooler atoms. We now estimate the initial temperature of the trap to evaluate the quality of the trap.

A simple estimation for the scattering rate would be to compare the difference in the number of counts with and without an atom in the focus. This simple estimation gives approximately 2500$s^{-1}$ as the scattering rate. Using a data collection time of 140ms, and recoil frequency of about 3.8kHz gives the energy increase as,
\begin{eqnarray}
  \Delta E &\approx& 2(2500s^{-1})(140ms)h(3.8 \textrm{kHz})
\end{eqnarray}
where the factor of 2 in the energy comes from considerations of both the absorption and emission processes.
Since the atom is in a harmonic potential, with a thermal density of the form \eqref{eqn:atom_density}, we require,
\begin{equation}
  \Delta E = \hbar \omega_{\rho} \Delta n_{\rho} + \hbar \omega_z \Delta n_z.
\end{equation}
The above equation can be solved numerically to give a temperature increase of, $\Delta T \approx 42 \mu\textrm{K}$. The temperature obtained from figure \ref{fig:ex vs T} is the averaged temperature over the probe time. Thus, assuming a linear increase of temperature, this gives an estimate of the initial temperature to be $T_{initial} \approx 160 \mu\textrm{K}$.

Although this temperature is above the Doppler cooling limit for $\Rb$, it should be noted that the atom is loaded from a magneto-optical trap into the dipole trap, which is $\approx 1 \textrm{mK}$ deep. During the loading of the atom into the trap, it is unavoidable that the atom heats up. Since this trap depth is an order of magnitude larger than the Doppler Cooling Limit, the predicted initial temperature is well within the expected temperature range.

We are thus confident that finite temperature effects of the atom accounts for the discrepancy reported in \cite{tey_interfacing_2009}. However, these predictions still have to be verified by a suitable temperature measurement, one possibility being that described in \cite{Tuchendler_energy_2008}, due to it's high similarity with the setup used. Another possibility would be to perform Raman sideband cooling, and inferring the temperature from there. The latter would seem the better choice, since the atom would be cooled in the process.

\section{Conclusion}
In this work, we present a further step in the characterization of the results of \cite{tey_strong_2008,tey_interfacing_2009}. We show that the inclusion of finite atomic temperature although intuitively simple, is not straightforward to implement, and numerical simulations seem like the only feasible option.
After we include this in the description of this trap, we compare theoretical predictions and experimental data, and show that the difference in the data can be attributed to the atomic temperature. After numerical studies, we obtain an effective atomic temperature of about $T_{initial} \approx 160 \mu\textrm{K}$ before probing of the atom . Although this temperature seems high, the discussion of section \ref{sec:meaning} gives us some insight as to the problem, and we are probably not even justified to make a comparison with the recoil or Doppler temperatures as found in the literature, due to a slightly differing definition of temperature used. Thus, we are unable to draw a definite temperature for the atom just from this study, and more work to measure and perhaps to cool the temperature is required.

\begin{acknowledgments}
  We would like to thank Wang Yimin, Syed Abdullah Aljunid, Gleb Maslennikov, Lee Jianwei and Christian Kurtsiefer for useful discussions. This work was supported by the National Research Foundation and the Ministry of Education, Singapore.
\end{acknowledgments}

\appendix

\section{Electric field}
\label{sec:appelec}
We set out here to give a more formal derivation for the Green's function used in equation \eqref{eqn:propagate}. This approach follows closely \cite{levine_theory_1950}. \\

For completeness, we repeat the following definitions. The electric field at any point is a solution of,
\begin{equation}
  \del \times (\del \times \vec{E}) -k^2\vec{E} = 0.  \label{eqn:vHelm_append}
\end{equation}
This equation can be solved with the help of the Dyadic Green's function defined as,
\begin{equation}
  \del \times (\del \times \dy{G}) -k^2\dy{G} = \dy{1} \delta(\vec{r}-\vec{r}\,'). \label{eqn:vHfG_append}
\end{equation}
Taking the divergence of \eqref{eqn:vHfG_append}, and substituting the resultant equation back, we obtain,
\begin{equation}
  -(\del^2+k^2)\dy{G} = (\dy{1} + \frac{\del\del}{k^2})\delta(\vec{r}-\vec{r}\,')
\end{equation}
This shows that if we have a Green's function which satisfies the scalar Helmholtz equation, ie.
\begin{equation}
  -(\del^2+k^2)g(\vec{r},\vec{r}\,') = \delta(\vec{r}-\vec{r}\,'), \label{eqn:sHg}
\end{equation}
the Green's function as defined in \eqref{eqn:vHfG_append} can be constructed as,
\begin{equation}
  \dy{G} =  (\dy{1} + \frac{\del\del}{k^2})g.
\end{equation}
We use the Green's function representing outgoing spherical waves,
\begin{equation}
  g(R) = \frac{e^{ikR}}{4\pi R} \label{eqn:g}
\end{equation}
where $R = \norm{\vec{r} - \vec{r} \, '}$.
To use the Green's function, we approach in a similar fashion to diffraction theory and consider the following difference,
\begin{equation}
  \vec{E}' \cdot (\del ' \times \del ' \times -k^2)\dy{G} - \big[(\del ' \times\del ' \times -k^2)\vec{E}'\big] \cdot \dy{G},  \label{vecform}
\end{equation}
where $\vec{E}' = \vec{E}(\vec{r}\,')$, and $\del '$ denotes differentiation with respect to the prime coordinates, and in the second term, we do not differentiate $\dy{G}$. Notice that we have preserved the order of the multipliction. Then, using Green's second vector identity, this becomes,
\begin{equation}
  \nabla'_i \bigg[\epsilon_{ijk}G_{j \lambda} \epsilon_{lmk}\nabla'_l E'_m - \epsilon_{ijk}E'_j\epsilon_{lmk}\nabla'_l G_{m \lambda} \bigg],
\end{equation}
where $\epsilon_{ijk}$ is the Levi-Civita tensor, and we assume the Einstein summation convention.
However, using equations \eqref{eqn:vHelm_append} and \eqref{eqn:vHfG_append}, equation \eqref{vecform} is nothing but, $\vec{E}' \delta(\vec{r} - \vec{r}\,')$. We thus have,
\begin{eqnarray}
  E'_\lambda \delta(\vec{r} - \vec{r}\,')  &=& \nabla'_i \bigg[\epsilon_{ijk}G_{j \lambda} \epsilon_{lmk}\nabla'_l E'_m -  {} \nn \\
  && {} \epsilon_{ijk}E'_j\epsilon_{lmk}\nabla'_l G_{m \lambda} \bigg]. \label{eqn:b4 int}
\end{eqnarray}
Since we only know the electric field on a plane, we need that the Dyadic Green's function be longitudinal on the plane, ie.
\begin{equation}
  \vec{n}' \times \dy{G}(\vec{r},\vec{r}\,')\Big|_{\vec{r} ' \in S} = 0. \label{eqn:BC}
\end{equation}
In our problem, the plane $S$, is the $z=0$ plane, and so $\vec{n} = \vec{e_z}$. Noticing that $\dy{G}(\vec{r},\vec{r}\,')$ is symmetric in both its arguments and that function resulting from operations on the right are still solutions of \eqref{eqn:vHfG_append}. Then, defining,
\begin{equation}
  \hat{U} = \dy{1} - 2  \vec{e_z}\vec{e_z},
\end{equation}
we can see that,
\begin{equation}
  \dy{G}^{(1)}(\vec{r},\vec{r}\,') = \dy{G}(\vec{r},\vec{r}\,') - \dy{G}(\vec{r},\hat{U}\vec{r}\,')\hat{U} \label{eqn:G 2 use}
\end{equation}
is a solution to \eqref{eqn:vHfG_append} satisfying the longitudinal condition \eqref{eqn:BC}. Taking the curl of \eqref{eqn:G 2 use}, substituting into \eqref{eqn:b4 int} and integrating, we obtain,
\begin{equation} \label{eqn:propagate_append}
  \vec{E}(\vec{r}) = \int_S dS' (ik)\frac{e^{ikR}}{2\pi R}
  \begin{pmatrix}
    -\frac{z}{R} & 0 & 0 \\
    0 & -\frac{z}{R} & 0 \\
    \frac{x'-x}{R} & \frac{y'-y}{R} & 0
  \end{pmatrix}
  \vec{E}(\vec{r}\,'),
\end{equation}
which is equation \eqref{eqn:propagate}.

\section{The form of the extinction}
\label{sec:app_ext}
The extinction is given in equation \eqref{eqn:epsilon} to be
\begin{equation}
  \epsilon = 1-{\Big \lvert 1+\frac{\inner{\vec{E}_{a}}{\vec{E}_{L}}}{\inner{\vec{E}_{L}}{\vec{E}_{L}}} \Big \lvert}^2.
\end{equation}
Firstly,
\begin{eqnarray}
  \inner{\vec{E}_{L}}{\vec{E}_{L}} &=& \iint d\rho'd\phi'\rho'(E_L e^{-\frac{\rho^2}{w_L^2}})^2 \nn \\
  &=& \frac{\pi E_L^2 w_L^2}{2}.
\end{eqnarray}
The next inner product is,
\begin{eqnarray}
  \inner{\vec{E}_{a}}{\vec{E}_{L}} &=& \int_{z=2f^+}dS \Big\{ \hat{\mathit{T}}\,' \vec{E}_a \Big\}^\dagger \vec{E}_L \nn \\
  &=& \frac{-3\pi E_L^2}{4} \tilde{I}^* (x_{a},z_{a}) \tilde{K} (x_{a},z_{a})
\end{eqnarray}
where $\tilde{I} (x_{a},z_{a})$ and $\tilde{K} (x_{a},z_{a})$ are,
\begin{widetext}
\begin{eqnarray}
    \tilde{I}(x,z) &=& \int^\infty_0 d\rho' \frac{\rho' z}{\sqrt{\cos\theta_{\rho'}}} (\cos\theta_{\rho'}+1) e^{-\frac{\rho^2}{w_L^2}} e^{-ik\sqrt{\rho'^2+f^2}}
  \int^{2\pi}_0\frac{d\phi'}{2\pi} \frac{e^{ikR}}{R^2} \label{eqn:I tilde def}
\end{eqnarray}
  \begin{eqnarray}
  \tilde{K}(x_{a},z_{a}) &=& \iint d\rho' \frac{d\phi'}{2\pi}\rho' \sqrt{\cos\theta_{\rho'}} \frac{e^{-\frac{\rho^2}{w_L^2}}}{r} e^{ik\sqrt{\rho'^2+f^2}-ikr} \times \Big(\cos\theta_{\rho'} + 1 + \frac{\rho'}{r^2}(\zeta\sin\theta_{\rho'}  {} \nn \\
  && {}-\cos\theta_{\rho'}(\rho'-x_a\cos\phi') +ix\sin\phi') -\frac{x_a}{2r^2}(\zeta\sin\theta_{\rho'}-\cos\theta_{\rho'}(\rho'-x_a\cos\phi') +ix_a\sin\phi')e^{-i\phi'} \Big) \label{eqn:Kappa def}
\end{eqnarray}
\end{widetext}
with
\begin{eqnarray}
  R &=& \sqrt{x_a^2+\rho'^2+z_a^2-2x_a\rho\cos\phi'}, \\
  \zeta &=& 2f-z_{a} \qquad \textrm{and}\\
  r &=&\sqrt{x_{a}^2+\zeta^2+\rho'^2-2\rho'\zeta\cos^2\phi'}.
\end{eqnarray}
Thus we have,
\begin{equation}
  \epsilon(x_a,z_a) = 1- \norm{1-\frac{3}{w_L^2}\tilde{I}^*(x_a,z_a) \tilde{K}(x_a,z_a) }^2.
\end{equation}


\begin{thebibliography}{99}

\bibitem{Haroche_book} S. Haroche, J.M. Raimond, \emph{Exploring the Quantum} (Oxford University Press, Oxford, 2006)

\bibitem{hennrich_transition_2005} M. Hennrich, A. Kuhn and G. Rempe, Phys. Rev. Lett. \textbf{94}, 053604 (2005)

\bibitem{mabuchi_cavity_2002} H. Mabuchi and A. C. Doherty, Science \textbf{298}, 1372 (2002)

\bibitem{ye_quantum_2008} J. Ye, H. J. Kimble and H. Katori, Scienc \textbf{320}, 1734 (2008)

\bibitem{ye_trapping_1999} J. Ye, D. W. Vernooy and H. J. Kimble, Phys. Rev. Lett. \textbf{83}, 4987 (1999)

\bibitem{armani_ultra_2003} D. K. Armani, T. J. Kippenberg, S. M. Spillane and K. J. Vahala, Nature \textbf{421}, 925 (2003)

\bibitem{kimble_quantum_2008} H. J. Kimble, Nature \textbf{453}, 1023 (2008)

\bibitem{M strong extinction} I. Gerhardt, G. Wrigge, P. Bushev, G. Zumofen, M. Agio, R. Pfab and V. Sandoghdar, Phys. Rev. Lett., \textbf{98}, 033601 (2007)

\bibitem{M2 strong extinction} G. Wrigge, I. Gerhardt, J. Hwang, G. Zumofen and V. Sandoghdar, Nature Physics \textbf{4}, 60 (2008)

\bibitem{tey_strong_2008} M. K. Tey, Z. Chen, S. A. Aljunid, B. Chng, F. Huber, G. Maslennikov and C. Kurtsiefer,
  Nature Physics, \textbf{4}, 924, (2008)

\bibitem{Syed strong 2009} S. A. Aljunid, M. K. Tey, B. Chng, T. Liew, G. Maslenikov, V. Scarani and C. Kurtsiefer, Phys. Rev. Lett. \textbf{103}, 153601 (2009)

\bibitem{van_enk_strongly_2001} S. J. van Enk and H. J. Kimble, Phys. Rev. A \textbf{63}, 023809 (2001)

\bibitem{tey_interfacing_2009}M. K. Tey, G. Maslennikov, T. C. H. Liew, S. A. Aljunid, F. Huber, B. Chng, Z. Chen, V. Scarani, and C. Kurtsiefer, New J. Phys. \textbf{11}, 043011 (2009).

\bibitem{Griffiths_book} D. J. Griffiths, \emph{Introduction to Electrodynamics}, (Prentice Hall, New Jersey, 3rd edition 1999)

\bibitem{Numerics: book frm paper} P. J. Davis and P. Rabinowitz, \emph{Methods of Numerical Integration} (Academic Press, Inc., Orlando, FL, 2nd Edition 1984)

\bibitem{Numerics: programming bible} W. H. Press, S. A. Teukolsky, W. T. Vetterling and B. P. Flannery, \emph{Numerical Recipes in C: The Art of Scientific Computing 2nd ed.} (Cambridge University Press, Cambridge, 1997)

\bibitem{huybreachs_evaluation_2007}{D. Huybrechs and S. Vandewalle, SIAM Jour. Numer. Analy. \textbf{44}, 1026 (2007)}

\bibitem{Tuchendler_energy_2008} C. Tuchendler, A. M. Lance, A. Browaeys, Y. R. P. Sortais and P. Grangier, Phys. Rev. A \textbf{78}, 033425 (2008)

\bibitem{levine_theory_1950}{H. Levine and J. Schwinger, Comm. Pure. Appl. Math \textbf{3}, 355 (1950)}

\bibitem{special func} M. Abramowitz and I. A. Stegun, \emph{Handbook of Mathematical Functions}, (Dover Publications, 1972)
\end{thebibliography}
\end{document}